\documentclass[12pt,a4paper]{article}
\usepackage[scale={.75,.8}]{geometry}
\usepackage{pstricks}
\usepackage{graphicx}
\usepackage[latin2]{inputenc}
\usepackage{epic}
\usepackage{latexsym}
\usepackage{epsf}
\usepackage{epsfig}
\usepackage{cite}
\usepackage{amssymb,amsmath}
\newcommand{\be}{\begin{equation}}
\newcommand{\ee}{\end{equation}}
\newcommand{\ba}{\begin{array}}
\newcommand{\ea}{\end{array}}
\begin{document}

\setcounter{page}{0}

\vspace{-1truecm}

\rightline{CERN-PH-TH-2006-223}
\rightline{CPHT-RR 082.1006}
\rightline{LPT--Orsay 06/68}
\rightline{hep-th/0610297}
\rightline{october 2006}

\vspace{1.cm}

\begin{center}

{\huge {\bf Moduli stabilization and uplifting \\ with dynamically generated F-terms }}
\vspace{1 cm}\\

{\large Emilian Dudas$^{1,2,3}$ ,  \ Chloe Papineau$^{3,2}$ and Stefan Pokorski $^4$
}
\vspace{1cm}\\

$^1$
CERN Theory Division, CH-1211, Geneva 23, Switzerland
\vspace{0.3cm}\\

$^2$
CPhT, Ecole Polytechnique 91128 Palaiseau Cedex, France
\vspace{0.3cm}\\

$^3$
Laboratoire de Physique Th\'eorique,
Universit\'e Paris-Sud, F-91405 Orsay, France
\vspace{0.3cm}\\

$^4$
Institute of Theoretical Physics,
Univ. of Warsaw, 00-681 Warsaw, Poland
\vspace{0.3cm}\\

\end{center}

\vspace{1cm}

\abstract{ We use the F-term dynamical supersymmetry breaking
models   with metastable vacua in order to uplift the vacuum energy
in the KKLT moduli stabilization scenario. The main advantage
compared to earlier proposals is the manifest supersymmetric
treatment and the natural coexistence of a TeV gravitino mass with a
zero cosmological constant. We argue that it is generically difficult
to avoid anti de-Sitter supersymmetric minima, however the tunneling
rate from the metastable vacuum with zero vacuum energy towards them can 
be very suppressed. We briefly comment on the properties of the
induced soft terms in the observable sector.}

\newpage

\vspace{3cm}

\newpage

\tableofcontents

\vspace{3cm}

%\newpage

%\setcounter{page}{1}
\pagestyle{plain}

\section{Introduction and Conclusions}

Chiral models of dynamical supersymmetry breaking with F-terms were constructed long time ago \cite{ads}.  
Explicit models with supersymmetry breaking ground state are generically relatively involved. 
More recently, Intriligator, Seiberg and Shih (ISS) proposed a simple,
vector-like model with long-lived, metastable supersymmetry breaking
vacua \cite{iss}, whereas the ground state is
supersymmetric\footnote{See \cite{iss2} for various extensions and
string embedding of the ISS proposal and \cite{ddgr} for an earlier proposal.}. 
On the other hand, in the last couple of years  convincing models of moduli stabilization in string theory were proposed, the propotype 
being the KKLT scenario \cite{kklt}, based on the orientifolds of IIB string theory flux compactifications \cite{gkp}.
One of the main problems of the KKLT scenario is the uplift of the vacuum
energy to zero or positive values. The original proposal of using
antibranes relies essentially on nonlinearly realized supersymmetry,
whereas the latter attempts \cite{Dudas:2005vv},\cite{dterms} to uplift vacuum
energy by D-terms, based on the suggestion in \cite{Burgess:2003ic},
lead generically to very heavy (close to the Planck mass) gravitino
mass\footnote{It would be very interesting to find explicit counter-examples to this claim. }.

Alternative uplifting using F-terms were already studied in \cite{silverstein,scrucca,lnr}.  
 As already stressed in \cite{scrucca}, \cite{lnr} and worked out in
detail in \cite{lnr}, a generic F-type supersymmetry breaking with a
supersymmetry breaking scale $ TeV \ll \Lambda_{SUSY} \ll M_P$ can
naturally produce the appropriate , intermediate energy scale,  for an uplift with a gravitino mass in the TeV 
range. Dynamical supersymmetry breaking 
is certainly the best candidate to fulfill this criterion. 
Metastable vacua have by definition a positive contribution to the vacuum energy which could clearly realize
the uplifting required in the KKLT scenario. As we will see in this
letter, dynamical supersymmetry breaking in metastable vacua of the
ISS type does achieve the goal of uplifting the KKLT vacuum energy
to zero, while keeping a TeV scale gravitino mass and therefore
leading to low energy supersymmetry. We would like to emphasize, however, that the main ingredient
in realizing the uplifting is not the metastable nature of the ISS model. 
Indeed, as we will briefly mention, other more traditional models \cite{it} of dynamical
supersymmetry breaking realize the uplifting in a qualitatively similar way. 
We argue by explicit examples in both cases that there are generically supersymmetric AdS minima
generated by the supergravity interactions, with however Planckian vev's for some fields and therefore
not fully trustable in the effective supergravity description. Even by considering seriously these AdS
minima, we argue that tunneling from the Minkowski metastable vacuum to the AdS supersymmetric one can be
very suppressed.

It would very interesting to couple the Minimal Supersymmetric
Standard Model to our present ISSKKLT setup, to work out the
low-energy phenomenology of the model and to compare it to the
existing works \cite{Choi:2004sx} based on the original KKLT
uplifting prescription relying on antibranes and nonlinearly
realized supersymmetry. 

The dynamically generated F-term uplifting method can also be
combined with the moduli stabilization in type IIA strings
\cite{IIA}. Indeed, D-term uplifing is not available in type IIA
strings with moduli stabilization, because of the strong constraints
coming from gauge invariance \cite{ibanez}. There are no such
constraints in our present setup, theferore there should be no
fundamental obstacles in uplifting vacuum energy by
nonsupersymmetric metastable vacua in type IIA strings with all
moduli stabilized . 

The structure of this note is as follows. In Section 2 we combine the KKLT model of moduli stabilization in
type IIB strings with the ISS model of metastable supersymmetry breaking vacuum. We show that in this
case the uplifting of the vacuum energy is naturally compatible with a TeV gravitino mass. We discuss supergravity
corrections to the globally supersymmetric vacuum, the possibility of a new supersymmetric minimum induced
by SUGRA interactions, the effects of gauging the color symmetry in the ISS model and the lifetime of the
metastable vacuum. In Section 3 we show  that qualitatively similar results are obtained by replacing
the ISS model with a more traditional model \cite{it} of dynamical supersymmetry breaking. In Section 4
we provide some general comments about the tree-level soft masses and under which conditions they could vanish. We then
apply the general formulae for the specific case of the model defined in Section 2 and work out some tree-level
soft terms, showing that generically tree-level soft masses are of the order of the gravitino mass, whereas
gaugino masses can be suppressed in particular cases.     
%%%%%%%%%%%%%%%%%%%%%%%%%%%%%%%%%%%%%%%%%%%%%%%%%%%%%%%%%%%%%%%%%%%%%%%%%%%%%%%%%%%%%%%%%%%%%%%%%%%%%%%%%%%%%%%%
\section{Metastable vacua and moduli stabilization}

The model is defined by
\begin{eqnarray}
&& W \ = \ W_1 (T) \ + \ W_2 (\chi^i) \ , \nonumber \\
&& K \ = \ - 3 \ \ln (T + T^{\dagger}) \ + \ |\varphi|^2 \ + \ |{\tilde \varphi}|^2 \ + \  |\Phi|^2 \ .
\label{iss1}
\end{eqnarray}
In (\ref{iss1})  $\chi^i$ denotes collectively the fields $\varphi_i^a$, ${\tilde \varphi}_a^{\bar j}$,
$\Phi_{\bar j}^i$ of the ISS model, where $i,{\bar j} = 1 \cdots N_f$ are flavor indices and $a,b = 1 \cdots N$ are colour indices.
Moreover, in (\ref{iss1})
\begin{eqnarray}
&& W_1 (T) \ = \ W_0 \ + \ a \ e^{-b T} \ , \nonumber \\
&& W_2 (\chi^i) \ = \ h \ Tr \ {\tilde \varphi} \ \Phi \ \varphi \ - \ h \ \mu^2 \ Tr \Phi \ . \label{iss2}
\end{eqnarray}
Notice that  the model is a staightforward combination of the ISS
model of metastable supersymmetry breaking vacua with the KKLT model
of moduli stabilization. As explained in \cite{iss}, the sector  $\varphi_i^a$, ${\tilde \varphi}_a^{\bar j}$ has a perturbative description
in the free magnetic range $N_f > 3 N$. 
The apropriate microscopic theory is an 
orientifold $IIB / \Omega'$ , with the orientifold operation
$\Omega' = \Omega (-1)^{F_L} I_6$, where $(-1)^{F_L}$ is the left
spacetime fermion number and $I_6$ is the parity in the six internal
coordinates. The theory contains D3 (O3) branes (orientifold planes) asked by the
orientifold operation, with the D3 branes placed at singular points of the compact space in order to reduce supersymmetry to ${\cal N}=1$. Typically there are
also D7 (O7) branes (orientifold planes) if other orbifold operations are present. The constant $W_0$ is
generated by 3-form closed string fluxes, as in \cite{gkp}, whereas
the nonperturbative $T$-dependent superpotential could come from
gaugino condensation on D7 branes \cite{kklt} or D3 brane
instantons. The gauge sector responsible for the nonperturbative ISS
dynamics has a natural embedding on a stack of $N$ D3 "color"
branes, with a dynamical scale depending on the dilaton field $S$, which was already stabilized by three-form fluxes. 
The mesonic fields $\Phi$ are naturally interpreted as
positions of a stack of $N_f$ D7 "flavor branes" . This could also
guarantees that their Kahler metric is independent at lowest order
on the volume Kahler modulus $T$, as already assumed in
(\ref{iss1}).  If the mesons would have entered into the no-scale structure of the T-modulus in (\ref{iss2}), as explained in \cite{scrucca}
the vacuum of the theory would have a marginally unstable direction. The quarks $\varphi$, ${\tilde \varphi}$ should come
from open string in the D3-D7 sector. We do not attempt here a
complete string construction underlying our effective theory, for
recent progress see \cite{iss2}. We point out nonetheless that
global string constructions with finite internal space volume are
needed in order to achieve this goal.

As transparent in (\ref{iss1}), the KKLT and the ISS sectors are
only coupled through gravitational interactions. In particular, as the ISS gauge group comes from D3 branes,
the dynamical scale in the electric theory and therefore also the mass parameter $\mu$ in the magnetic theory
superpotential (\ref{iss2}) depend on the dilaton $S$, which we assume is already stabilized by NS-NS and RR 
three-form fluxes. We believe this decoupling is instrumental in getting the uplift of the vacuum energy. 
Another reason for forbidding a coupling to the $T$ modulus of the
dynamical supersymmetry breaking sector in the global supersymmetric
limit is that it is unclear how to formulate the nonabelian Seiberg
duality for field-dependent couplings.

At the global supersymmetry level and before gauging the color
symmetry, the ISS model has a global symmetry $G = SU(N) \times
SU(N_f) \times SU(N_f) \times U(1)_B \times U(1)' \times U(1)_R$,
broken explicitly to $ SU(N) \times SU(N_f) \times U(1)_B \times
U(1)_R$ by the mass parameter $\mu$. In the supergravity embedding
(\ref{iss2}), the R-symmetry $U(1)_R$ is explicitly broken.
 To start with, we consider the ungauged theory, in which
the $SU(N)$ is part of the global symmetry group. At the global
supersymmetry level, the metastable ISS vacuum is
\begin{equation}
\Phi_0 \ = \ 0 \quad , \quad  \varphi_0 \ = \ {\tilde \varphi}_0^T \ =
 \left(
\begin{array}{c}
\mu I_N \\
0
\end{array}
\right)
 \ , \label{iss3}
\end{equation}
where $I_N$ is the $N \times N$ identity matrix and $\mu \ll
\Lambda_m$, where $\Lambda_m \le M_P$ denotes the mass scale
associated with the Landau pole for the gauge coupling in the
magnetic theory. The first question to address is the vacuum
structure of the model. In order to answer this question, we start
from the supergravity scalar potential 
\begin{equation}
V \ = \ e^{K}  \left[ (K^{-1})^{i {\bar j}}  D_i W  D_{\bar j}  {\bar W}  \ - \  3  |W|^2   \right]  \ + \
{1 \over 2} \ (Re f_a) \ D_a^2 \ , \label{vsugra}  
\end{equation}
where $Re f_a = 1/g_a^2$ define the gauge couplings . By using\footnote{The gauge D-term contributions do not exist 
in the un-gauged case we are
discussing in this section and will play essentially no role in the following sections.} 
(\ref{iss1})-(\ref{iss2}), we find 
\begin{equation}
V \ = \ {e^{{\bar \chi}_{\bar i} \chi^i} \over (T + {\bar T})^3} \
\{  {(T + {\bar T})^2 \over 3} |\partial_T W - {3 \over T + {\bar T}} W |^2 +
\sum_i | \partial_i W \ + \ {\bar \chi }_{\bar i} W |^2 \ - \ 3 |W|^2  \} \ . \label{iss4}
\end{equation}

Since $\mu \ll M_P, $ the vev's in the ISS model are well below the Planck scale. Then an illuminating way of rewriting
the scalar
potential (\ref{iss4}) is to expand it in powers of the fields $\chi^i/M_P$, in which case it reads\footnote{In most
of the formulae of this letter, $M_P=1$. In some formulae, however, we keep explicitly $M_P$.}
\begin{eqnarray}
&& V \ = \ {1 \over  (T + {\bar T})^3} \ V_{ISS} (\chi^i, {\bar \chi}_{\bar i}) \ + \ V_{KKLT} (T,{\bar T}) \ + \
{ {\bar \chi}_{\bar i}  \chi^i \over M_P^2} \ V_1  (T,{\bar T}) \ \nonumber \\
&& + \ {1 \over M_P^3} \ \left[ \ W_2 (\chi^i) \ V_2 (T,{\bar T}) + \chi^i \partial_i W_2 \ V_3 (T,{\bar T}) \ +
\ h.c. \right] \ + \ \cdots \ , \label{iss5}
\end{eqnarray}
where by comparing (\ref{iss5}) with (\ref{iss4}) we can check that
$V_1 \sim m_{3/2}^2 M_P^2$, $V_2, V_3 \sim m_{3/2} M_P^3$, where as
usual $m_{3/2}^2 = |W^2| \exp (K)$.  
Notice that the contribution to the vacuum energy from the ISS sector, in the global limit, is
\begin{equation}
\langle V_{ISS} \rangle \ = \ (N_f-N) \ h^2  \ \mu^4 \ . \label{iss05} 
\end{equation}
Since
we are interested
in small (TeV scale) gravitino mass, it is clear that the first two terms in the rhs of (\ref{iss5}), $V_{ISS}$ and
$V_{KKLT}$ are the leading terms. Consequently, there should be a vacuum very close to a uplift KKLT vacuum $\langle T \rangle = T_0$
and the ISS vacuum $\langle \chi^i \rangle = \chi^i_0 $.  The KKLT uplift vacuum at the zeroth order $T_0$ is defined as
the minimum of the zeroth order potential $\partial_{T_0} V_0 =0 $, obtained by inserting the ISS vacuum (\ref{iss3})
into the supergravity scalar potential
\begin{equation}
V_0 \ = \ {1 \over (T + {\bar T})^3} \ \left[ {(T+{\bar T})^2 \over 3} |D_T W_1|^2 - 3 |W_1|^2 + h^2  (N_f-N) \mu^4 \right] \
\ . \label{iss06}
\end{equation}
In the limit $b T \gg 1$ and for zero cosmological constant, a good approximation for $T_0$, considered to be real in what 
follows, is provided by
\begin{equation}
W_0 \ + \ {a b (T_0+{\bar T}_0) \over 3} \ e^{ - b T_0} \ = \ 0 . \label{iss07}
\end{equation}
Notice that in this case $T$ does contribute to supersymmetry breaking\footnote{Notice that the leading order expression for
$W_0$ in (\ref{iss07}) is not enough for computing $F^T$, since the subleading terms neglected in (\ref{iss07}) are needed as well.
$F^T$ can be computed directly, however, by keeping the leading terms in the eq. $\partial_T V =0$.}
\begin{equation}
F^T \ \equiv \  e^{K \over 2} \ K^{T {\bar T}} \ \overline{D_T W} \ \simeq \
 \ {a \over (T_0 + {\bar T}_0)^{1 /2}} \ e^{-b T_0} \
, \label{iss08}
\end{equation}
 but by an amount supressed by a factor of
$1/ b (T_0 + {\bar T}_0)$ compared to the naive expectation. 

The cosmological constant at the lowest order is  given by
\begin{equation}
\Lambda \ = \ V_{KKLT} (T_0, {\bar T}_0) \ + \ {(N_f-N) h^2 \mu^4 \over (T_0 + {\bar T}_0)^3} \ , \label{iss7}
\end{equation}
which shows that the ISS sector plays the role of un uplifting sector of the KKLT model. In the zeroth order approximation
and in the large volume limit $b (T_0 + {\bar T}_0) \gg 1 $, we find that the condition of zero cosmological constant
$\Lambda = 0$ implies roughly
\begin{equation}
3 \ |W_0|^2 \ \sim \ h^2 \ (N_f-N) \ \mu^4 \ . \label{iss8}
\end{equation}
If we want to have a gravitino mass $m_{3/2} = \sim
W_0 / (T_0 + {\bar T}_0)^{3/2}$ in the TeV range, we need small
values of $\mu \sim 10^{-6} - 10^{-7}$. Since $\mu$ in the model of
\cite{iss} has a dynamical origin, this is natural. Moreover, the
metastable vacuum of \cite{iss} has a significantly large lifetime
exactly in this limit, more precisely when $\epsilon \equiv (\mu /
\Lambda_m) \ll 1$. Therefore, a light (TeV range) gravitino mass is
natural in our model and compatible with the uplift of the
cosmological constant. We believe that this fact is an 
improvement over the D-term uplift models suggested in
\cite{Burgess:2003ic} and worked out in \cite{dterms}.

Notice that supergravity corrections give tree-level masses to the
pseudo-moduli fields of the ISS model. As explained in more general
terms in \cite{iss}, these corrections are subleading with respect
to masses arising from the one-loop Coleman-Weinberg effective
potential in the global supersymmetric limit. This can be explicitly
checked starting from the supergravity scalar potential (\ref{iss4})
and expanding in small fluctuations around the vacuum (\ref{iss3})
to the quadratic order.
%%%%%%%%%%%%%%%%%%%%%%%%%%%%%%%%%%%%%%%%%%%%%%%%%%%%%%%%%%%%%%%%%%%%%%%%%
\subsection{The metastable vacuum and supergravity corrections}

By coupling the T field to the ISS dynamical supersymmetry breaking system, we expect small deviations from the 
lowest order vacuum  (\ref{iss3}), (\ref{iss07}). We expand
\begin{equation}
\chi^i \ = \ \chi^i_0 +\delta \chi^i \quad , \quad T \ = \ T_0 \ + \ \delta T \ , \label{iss6}
\end{equation}
where $\chi_0^i$ are provided by (\ref{iss3}), with $\delta \varphi  \ll  \varphi_0$
( $ \delta {\tilde \varphi}  \ll {\tilde \varphi}_0$) and  $\delta T \ll T_0$.
We now turn to the SUGRA corrections to the ISS metastable vacuum (\ref{iss6}), by
linearizing around the KKLT-ISS vacuum the field eqs,
\begin{equation}
\partial_{\varphi} V \ = \ \partial_{\tilde \varphi} V \ = \  \partial_{\Phi} V \ =  \ \partial_T V \ = \ 0 \ , \label{corr1}
\end{equation}
This can be done by starting from the expansion in the fields $\chi$ in (\ref{iss5}), where
\begin{eqnarray}
&& V_1 \ = \ V_{KKLT} \ + \ { |W|^2 \over (T + {\bar T})^3} \ , \label{corr2}  \\
&&  V_2 \ = \ - 
 { 1 \over (T + {\bar T})^3} \left[ (T + {\bar T}) \ \overline{D_T \ W} \ - \ 3 \ \overline{W_1} \right] \quad , \quad
  V_3 \ = \ { {\overline{W_1} \over (T + {\bar T})^3}} \ . \nonumber 
\end{eqnarray}
Notice that in the zeroth order vacuum $V_1 \sim m_{3/2}^2 M_P^2$, $V_2,V_3 \sim m_{3/2} M_P^3 $, as well as
$\partial_T V_1 \sim m_{3/2}^2 M_P^2$ and $\partial_T V_2, \partial_T V_3 \sim m_{3/2} M_P^3 $. 
In order for the linearization to be well-defined, we need to include the Coleman-Weinberg one-loop quantum corrections
to the scalar potential discussed in \cite{iss}. The reason is that at tree-level and in our zeroth order approximation,
there are zero mass particles which, in addition to the Goldstone bosons of the broken symmetries, contain also pseudo-moduli
which get their masses at one-loop. After including these corrections,
we find at the leading order in the variations $\delta \chi^i, \delta T$ and for zero cosmological constant, that
\begin{equation}
\delta \chi^i \ \leq \ O(m_{3/2}) \qquad , \qquad \delta T \ \leq O({m_{3/2} \over M_P}) \ . \label{corr3} 
\end{equation}
Since in our framework $m_{3/2} \ll \mu$, the
condition $\delta \varphi \ll \varphi_0 $ is largely satisfied, showing
that the expansion (\ref{iss6}) is an excellent approximation. The precise values of the supergravity corrections (\ref{corr3}) are not 
important for what follows. Notice that the small values for $\delta \varphi$, $\delta \Phi$ in (\ref{corr3}) 
are in agreement with the arguments given in \cite{iss} stating that  high energy microscopic effects in the magnetic theory
should not affect significantly the metastable vacuum.   

%%%%%%%%%%%%%%%%%%%%%%%%%%%%%%%%%%%%%%%%%%%%%%%%%%%%%%%%%%%%%%%%%%%%%%%%%%%%%%%%%
\subsection{The SUGRA induced magnetic supersymmetric minimum}

In the ISS model and in the case of ungauged $SU(N)$ symmetry, the ISS vacuum (\ref{iss3})
is actually the true ground state. What happens in the supergravity embedding we are proposing here ?
We will show that there is a new, AdS supersymmetric ground state generated by the SUGRA interactions.
To find it, we search solutions of the type
\begin{eqnarray}
&&  \varphi \ \ = \  \ \left(
\begin{array}{c}
\varphi_1 \\
0
\end{array}
\right) \quad , \quad
 {\tilde \varphi}^T \  \ = \ \ \left(
\begin{array}{c}
{\tilde \varphi}_1 \\
0
\end{array}
\right) \ ,
  \nonumber \\
&& \Phi \ = \
\left(
\begin{array}{cc}
\Phi_1  & 0
\\
0 & \Phi_2
\end{array}
\right) \ , \label{susy1}
\end{eqnarray}

of the SUSY preserving equations
\begin{eqnarray}
&& D_{\varphi} W \ = \ 0 \quad \rightarrow \quad h \ {\tilde \varphi}_1 \Phi_1 + \overline{\varphi}_1 \ W \ = \ 0 \ ,
\label{susy2} \\
&& D_{\tilde \varphi} W \ = \ 0 \quad \rightarrow \quad
h \ \Phi_1 {\varphi}_1 + \overline{\tilde \varphi}_1 \ W \ = \ 0 \ , \nonumber \\
&& D_{\Phi} W \ = \ 0 \quad \rightarrow \quad  h \left(  {\tilde \varphi}_1^i  \varphi_{1,j} - \mu^2 \delta_j^i \right) \ +
\ ({\bar \Phi}_1)_j^i \ W \ = \
0 \ , \ i,j \ = \ 1 \cdots N \nonumber \\
&& D_{\Phi} W \ = \ 0 \quad \rightarrow \quad  - h \ \mu^2 \delta_m^n \ + \ ({\bar \Phi}_2)_m^n \ W \ = \ 0 \ , \ m,n \ =
\ N+1 \cdots N_f \ ,  \nonumber \\
&& D_T \ W \ = \ 0  \quad \rightarrow \quad  a \ b \ e^{-b T_m} \  + \ {3 \over T_m + {\bar T}_m} \ W \ = \ 0 . \nonumber
\end{eqnarray}
The eqs. (\ref{susy2}) have the following solution :
\begin{eqnarray}
&& \varphi_1 \ = \ \mu_1 \ I_N \quad , \quad  {\tilde \varphi}_1 \ = \ \mu_2 \ I_N \quad ,
\quad {\rm with} \ \ |\mu_1| \ = \ |\mu_2| \ , \nonumber \\
&& \Phi_1 \ = \  (\mu_1 \mu_2 - \mu^2)^{1 \over 2} \ I_N \quad , \quad \Phi_2 \ = \ - \ {\mu^2 \over (\mu_1 \mu_2 - \mu^2)^{1 \over 2}} \
I_{N_f-N} \ , \nonumber \\
&& a \ b \ e^{- b T_m} \ - \ {3 h \over T_m + {\bar T}_m} \  (\mu_1 \mu_2 - \mu^2)^{1 \over 2} \ = \ 0 \ , \nonumber \\
&& h^2 \ (\mu_1 \mu_2 \ - \ \mu^2) \ - \ |W|^2 \ = \ 0 \ . \label{susy3}
\end{eqnarray}
Since cosmological constant cancellation asks for $m_{3/2} \sim
\langle W \rangle \sim h \mu^2$, where $m_{3/2}$ is the gravitino
mass in the ISS-KKLT vacuum, for $\mu_i \sim \mu$ eq. (\ref{susy3})
implies in particular $\Phi_2 \sim M_P$, the supersymmetric minimum
(\ref{susy3}) depends on the UV properties of the model and is not
fully reliable in our effective field theory analysis. For $\mu_1 \mu_2 \gg \mu^2$, all
vev's are well below $M_P$, $\langle W \rangle \gg m_{3/2} M_P^2$
and the supersymmetric vacuum (\ref{susy3}) would be within the
validity of the effective supergravity. The second possibility is
however incompatible with the condition (\ref{iss8}) and for a TeV
gravitino mass. Therefore we recover the conclusion that $\Phi_2 \sim M_P$.

 Notice that the supersymmetric vacuum (\ref{susy3}) survives the gauging of the $SU(N)$
 symmetry. Indeed, the $SU(N)$ D-flatness conditions are satisfied, since $|\varphi_1|^2 = |\varphi_2|^2 $ 
and $ [\Phi , \Phi] = 0 $ in (\ref{susy3}).
%%%%%%%%%%%%%%%%%%%%%%%%%%%%%%%%%%%%%%%%%%%%%%%%%%%%%%%%%%%%%%%%%%%%%%%%%%%%%%%%%%%%%%%%%%%%%%%%%%%%%%%%5
\subsection{Gauging the model : infrared description}

In the ISS model, the $SU(N)$ symmetry is gauged and corresponds to the gauge group of the magnetic theory. 
In the electric description, the ISS model is the supersymmetric QCD with $N_c$ colors and 
$  N_c < N_f  <  3N_c /2$ quark flavors $Q. {\tilde Q}$, such that in the magnetic  description with the gauge group
$SU(N_f-N_c)$,
the number of flavors  is large $N_f > 3 N$, where the magnetic theory is in the infrared-free phase.  In this case the perturbative
magnetic description, around the origin in field space, is reliable.  The electric theory has a dynamical scale
$\Lambda$ and a  mass term for the quarks $W = m_i^{\bar j} Q^i  {\tilde Q}_{\bar j}$. There are $N_c$  vacua described by
\begin{equation}
M_{\bar j}^i  \ = \  ({1 \over m})_{\bar j}^i  \ (det m)^{1 \over N_c} \ \Lambda^{3N_c-N_f \over N_c} \ . \label{gauge01}
\end{equation}
The perturbative treatment in the magnetic description translates into the constraint $m_a \ll  \Lambda $, where
$a$ denotes here the number of light mass eigenvalues, which has to be equal or larger to $N_f+1$ in order
for the metastable vacua to exist.  One of the open questions for the ISS model is a dynamical explanation for the
constraint $m_a \ll \Lambda $. We believe that a simple possibility is the following. At high energy there is an additional
abelian " anomalous " symmetry $U(1)_X$, with mixed anomalies $U(1)_X SU(N_c)^2$ cancelled by the Green-Schwarz
mechanism involving an axionic field $a_X$.    This will render the gauge vector $V_X$ massive and stabilize the 
complex modulus field containing the axion $a_x$ . There will be an induced Fayet-Iliopoulos
term, which in explicit string models is always cancelled by the vev of a scalar field $\langle N \rangle \ll M_P$.  
 Mixed anomalies mean that the sum of
charges quark charges $X_Q + X_{\bar Q}$ is not zero and therefore the  mass operator $m_i^{\bar j} Q^i  {\tilde Q}_{\bar j}$ is not gauge invariant.  
In generic models, the charge $X_N$ is oppposite compared to   $X_Q + X_{\bar Q}$. We normalize $X_N=-1$ in what follows.
Then the superpotential term  
$y_i^{\bar j} (N/M_P)^{X_Q + X_{\bar Q}} Q^i  {\tilde Q}_{\bar j}$ is perturbatively allowed. 
Supersymmetry  could be broken in the process \cite{bd},
but it can also stay unbroken. In this last case, at energy scales well below the mass of the gauge boson $A_X$, the net effect
of all this is to generate an effective mass term for the quarks of the electric theory $m \sim  (\langle N  \rangle / M_P)^{X_Q + X_{\bar Q}} $.
For large enough quark charges and/or small enough vev $\langle N \rangle$, the induced mass $m$ can be 
very small. Another generical way of getting small masses was proposed recently in \cite{dfs}.    

 Denoting by $\Lambda_m$ the Landau pole of the magnetic theory, according to ISS for arbitrary vev's of $\Phi$ the quark flavors
become massive and can be integrated out. By doing this and by coupling the resulting low-energy system to the
KKLT model, we arrive at a lagrangian described by
\begin{eqnarray}
&& W \ = \ W_0 \ + \ a \ e^{-b T} \ + \ N \ \left( {h^{N_f} {det \Phi} \over \Lambda_m^{N_f - 3 N}} \right)^{1/N}
\ - \ h \ \mu^2 \ Tr \Phi \ , \nonumber \\
&& K \ = \ - 3 \ \ln (  T + {\bar T}) \ + \ {\bar \Phi} \Phi \ . \label{gauge1}
\end{eqnarray}
Similarly to the global supersymmetry analysis of ISS \cite{iss}, this action has $N_f-N$ supersymmetric vacua,
which in the global limit are given by
\begin{equation}
\langle h \Phi \rangle \ = \ \Lambda_m \epsilon^{2 N / (N_f-N)} \ I_{N_f} =
\ \mu \ {1 \over \epsilon^{(N_f-3 N)/(N_f-N)}} \ I_{N_f} \ , \label{gauge2}
\end{equation}
where $\epsilon \equiv \mu / \Lambda_m$. The vacuum in the T-direction is simpler to describe  by replacing the
vev's (\ref{gauge2}) in the superpotential (\ref{gauge1}). By doing this, we get an effective superpotential
\begin{equation}
W_{\rm eff} \ = \ W_0 \ - \ {(N_f-N) \mu^3 \over \epsilon^{(N_f-3N)/(N_f-N)}} \ + \ a \ e^{-b T} \ .
\label{gauge3}
\end{equation}
Since $W_0 < 0$ in the KKLT model, the effect of the supersymmetric $\Phi$ vev's is to increase
the absolute value
of the (negative) constant in the superpotential. The approximate values of the minimum for $T$ and the
corresponding negative cosmological constant are given approximately by
\begin{eqnarray}
&& a \ b \ e^{-b T_s} \ + \ {3 \over T_s + {\bar T}_s} \left( W_0 \ -{(N_f-N) \mu^3 \over \epsilon^{(N_f-3N)/(N_f-N)}}
\ \right) \ \simeq \ 0 \ , \nonumber \\
&& V_0 \ \simeq \ - { 3 \over (T_s + {\bar T}_s)^3}   | W_0 \ -{(N_f-N) \mu^3 \over \epsilon^{(N_f-3N)/(N_f-N)}} |^2
\ . \label{gauge4}
\end{eqnarray}
The supersymmetric ISS vacuum is therefore AdS . Notice that for
$W_0 \gg \mu^3 / \epsilon^{(N_f-3 N)/(N_f-N)}$, we get $T_s \sim
T_0$, with $T_0$ defined in (\ref{iss07}), since in this case $W \simeq W_0$. If $W_0 \ll \mu^3 / \epsilon^{(N_f-3 N)/(N_f-N)}$,
then $T_s  < T_0$.

%%%%%%%%%%%%%%%%%%%%%%%%%%%%%%%%%%%%%%%%%%%%%%%%%%%%%%%%%%%%%%%%%%%%%%%%%%%%%%%%%%%%%%%%%%%%%%%%%%%%%%%%%%%%%%%%%%%%%%%%
\subsection{Lifetime of the metastable vacuum}

The model we discussed in this paper has one metastable vacuum and two type of AdS supersymmetric minima.
The metastable vacuum will tunnel to the supersymmetric AdS minimum (\ref{gauge2})-(\ref{gauge4}). 
The purpose of this section
is to provide a qualitative estimate of the lifetime of the metastable minimum, following \cite{coleman},\cite{duncan}.
  The bounce action is expected to come from the path in field space of minimum potential barrier between the metastable supersymmetry
breaking vacuum and the supersymmetric vacua. Along this path, the bounce action cannot be computed analytically. For
a triangular idealized approximation \cite{duncan}, the bounce action  $S_b$ is qualitatively
\begin{equation}
S_b \ \sim \ {(\Delta \chi)^4 \over \Delta V} \ , \label{tunneling1}
\end{equation}
where $\Delta V$ is the (minimum) barrier along the bounce and $\Delta \chi$ is the variation of the relevant field.
For the tunneling between the metastable ISS vacuum (\ref{iss3}) and the supersymmetric one (\ref{gauge2}) after
gauging $SU(N)$, there are two cases. If $\mu \ll \epsilon^{(N_f-3 N)/(N_f-N)} M_P$, we get
\begin{equation}
h \ \Delta \Phi \ \simeq \ \mu \ {1 \over \epsilon^{(N_f-3 N)/(N_f-N)}} \quad , \quad \Delta V \ \sim \
  { 3 \over (T_s + {\bar T}_s)^3} \ |W_0|^2 \ . \label{tunneling2}
\end{equation}
Then, by using the condition (\ref{iss8}) of the vanishing of the
vacuum energy in the metastable vacuum , we get
\begin{equation}
S_b \ \sim \ {  (T_s + {\bar T}_s)^3 \over \epsilon^{4 (N_f-3 N)/(N_f-N)}} \ \gg 1 \ , \label{tunneling3}
\end{equation}
which increases the lifetime of the metastable vacuum compared to the similar ISS analysis.
The reason is that the energy difference between the metastable and
the AdS supersymmetric minimum is decreased by the factor $1 / (T_s
+ {\bar T}_s)^3 $, resulting in an increase in the bounce action
$S_b$.
In the case where $\mu \gg \epsilon^{(N_f-3 N)/(N_f-N)} M_P$, the vacuum energy of the supersymmetric vacuum (\ref{gauge4})
and consequently $\Delta V$ change. The bounce action in this case is
\begin{equation}
S_b \ \sim \  {M_P^2 \over \mu^2}  {  (T_s + {\bar T}_s)^3 \over \epsilon^{2 (N_f-3 N)/(N_f-N)}} \ \gg 1 \ . \label{tunneling4}
\end{equation}
The metastable minimum could also tunnel to the supersymmetric
minimum (\ref{susy3}). Even by taking seriously the effective theory
analysis in this case,  we notice that the AdS supersymmetric minimum (\ref{susy3}) is
far away in the $\Phi$ field space from the ISS-KKLT metastable
vacuum (\ref{iss3}), (\ref{iss07}). The tunneling probability to go
to the AdS vacuum (\ref{susy3}) is highly suppressed and irrelevant
for all practical purposes.
%%%%%%%%%%%%%%%%%%%%%%%%%%%%%%%%%%%%%%%%%%%%%%%%%%%%%%%%%%%%%%%%%%%%%%%%%%%%%%%%%%
\section{Uplifting with supersymmetry breaking on the quantum moduli space}

As mentioned in the introduction,  the important ingredient from the F-term dynamical supersymmetry breaking sector is the intermediate scale
for the resulting (positive) contribution to the vacuum energy and not the metastable nature of the vacuum.  We discuss now  a more conventional
non-perturbative hidden sector which, in the global supersymmetry limit, has a non-supersymmetric ground state \cite{it}. Since most
of the analysis parallels that already done for the ISS model, our discussion will be very brief.  We consider a SQCD model with $N_c=N_f=2$ 
colors and flavors. The effective action which puts together 
the KKLT moduli stabilization sector and the supersymmetry breaking sector is
\begin{eqnarray}
&& W \ = \ W_0 \ + \ a \ e^{-b T}  \ + \  \lambda S^{ij} M_{ij}  \ + \ X \ (Pf M - \Lambda_2^4)  \  ,  \nonumber \\
&& K \ = \ - 3  \ \ln (T + {\bar T})  \ + \  Tr ( {1 \over \Lambda_2^2} \ |M|^2 \ + \ |S|^2)    \ , \label{it1}
\end{eqnarray}
where $Pf M \ = \ \epsilon^{ijkl} M_{ij} M_{kl}$,  $\Lambda_2$ is the dynamical scale of the theory,  $M_{ij} = Q_i^a Q_j^a$ are the mesons builded up from
the quarks $Q_i^a$ with color indices $a=1,2$ and flavor indices $i,j=1,2,3,4$, whereas  $S^{ij}$ are gauge singlets. Both fields are antisymmetric in the 
flavor indices. In (\ref{it1}), $X$ is a lagrange multiplier which enforces 
the  eq. describing the quantum deformed moduli space $Pf M = \Lambda_2^4$, whereas the factor of  $(1 /  \Lambda_2^2)$ in the Kahler potential
of the mesons is present since mesons have mass dimension two and have a dynamical origin.  The supergravity scalar potential resulting from (\ref{it1}) is
\begin{eqnarray}
&& V \ = \ { e^{  Tr ( (|M|^2 / \Lambda_2^2)  + \ |S|^2)   } \over (T + {\bar T})^3} \
\{  {(T + {\bar T})^2 \over 3} |\partial_T W - {3 \over T + {\bar T}} W |^2 \ + \ \sum_{ij}  |\lambda M_{ij} + {\bar S}_{ij} W|^2 \nonumber \\
&& 
+ \ \sum_{ij}  |  \lambda S^{ij} + 2 X \epsilon^{ijkl} M_{kl} + {{\bar M}^{ij} \over \Lambda_2^2}   W |^2 \ + 
\ |Pf M - \Lambda_2^4|^2 \ - \    \ 3 |W|^2  \} \ . \label{it2}
\end{eqnarray}
In the global limit, the strongly coupled sector break supersymmetry, since there is no solution to the supersymmetry eqs. 
$F^X = F^S=0$. As  explained in \cite{it}, the strongly coupled sector produces a contribution to the vacuum energy of the order
\begin{equation}
V_0 \ \sim \ \lambda^2 \Lambda_2^4 \ . \ \label{it3}
\end{equation}
Even if at the global supersymmetric level, the ground state breaks supersymmetry, similarly to the ISS model discussed in section 2.2,
at the supergravity level we do find a supersymmetric AdS minimum.   Indeed, by inserting the maximally, $SO(5)$ symmetric ansatz 
\begin{equation}
 \langle M  \rangle \ = \
\left(
\begin{array}{cc}
i \sigma_2  & 0
\\
0 & i \sigma_2 
\end{array}
\right) \  \Lambda_2^2 ,  \qquad , \qquad 
\langle S  \rangle \ = \ c \ 
\left(
\begin{array}{cc}
i \sigma_2  & 0
\\
0 & i \sigma_2
\end{array}
\right) \  \Lambda_2^2 \ ,  
\label{it4}
\end{equation}
into the supersymmetry conditions $D_S W = D_M W = D_X W = D_T W =0$, we find
\begin{eqnarray}
&& \lambda \ + \ c \ W \ = \ 0 \qquad , \qquad \lambda \ c \ + \ 2 \ X \ + \ {W \over \Lambda_2^2} \ = \ 0 \ , \nonumber \\
&& a \ b \ e^{-b T_0}  \ + \ {3 \over  T_0 + {\bar T}_0} \  \left( W_0 \ + \ a \ e^{-b T_0} \ + \ 4  \ \lambda \ c \ \Lambda_2^4    \right) \ = 0 \ . \label{it5}  
\end{eqnarray}   
If these conditions have a solution, the original supersymmetry breaking ground state becomes metastable. The
condition for the uplifting of the vacuum energy in the metastable
vacuum requires then $W_0 \sim \lambda \Lambda_2^2$. 
The last eq. in
(\ref{it5}) leads then, for $b T_0 \gg 1$, to $W \sim W_0$ in a first
approximation, whereas $T_0$ is given again by (\ref{iss07}). TeV
values for the gravitino mass asks therefore for $\Lambda_2^2 \sim
m_{3/2} M_P \sim (10^{11} \ GeV)^2$. 
Combining
the first two eqs. in (\ref{it5}), we then find $c \sim - \lambda / W_0$ and therefore $\langle S \rangle  \sim M_P$.  We find therefore, analogously to section 2.2,
Planckian values for the supersymmetric AdS vacuum, which signifies that the supersymmetry preserving vacuum is actually beyond the
regime of validity of the effective lagrangian description. In contrast to section 2.2, however, the AdS vacuum energy itself is Planckian here
$V_{AdS} \sim \lambda^2 M_P^4$.  
 
 By taking seriously this supersymmetric solution, the tunneling from the non-supersymmetric metastable vacuum proceed in the
 S-field direction in the field space. Since $\Delta S \sim M_P $, whereas $\Delta V = |V_{AdS}| \sim \lambda^2 M_P^4$, we find for the
 bounce action $S_b \sim (1 /  \lambda^2 )$. The tunneling probability  $\exp(-S_b) $ is therefore suppressed only in the $\lambda \ll 1$ limit.
 This condition is the analog of the condition $m \ll \Lambda$ in the
 electric version of the ISS model , i.e.  the quarks  must have
 masses much smaller
than the dynamical scale of the electric theory.

%%%%%%%%%%%%%%%%%%%%%%%%%%%%%%%%%%%%%%%%%%%%%%%%%%%%%%%%%%%%%%%%%%%
\section{Soft terms and mass scales }

\subsection{General tree-level formulae }

The relevant couplings for our present
discussion are the following terms in the Kahler potential and the superpotential arising in the perturbative expansion 
in the matter fields $M^I$
\begin{eqnarray}
&& K \ \rightarrow \ K \ + \  \left[ (T + {\bar T})^{n_I} \ Z_{I {\bar J}} + \cdots \right] \ \ M^I  {\bar M}^{\bar J} \ + \
\cdots \   \equiv  K  +   K_{I  {\bar J}}  M^I {\bar M}^{\bar J}  \  , \nonumber \\
&& W  \  \rightarrow \ W \ + \ {1 \over 6} \ W_{IJK} \ M^I \ M^J \ M^K
 \ , \label{general01}
\end{eqnarray}
where $\cdots$ denote couplings to other  (hidden-sector, messengers in gauge mediation models, etc) fields. 
 In a manifestly supersymmetric approach, with both F and D-term
contributions, the condition of zero cosmological constant is 
\begin{equation}
K_{\alpha \bar \beta}  F^{\alpha}   F^{\bar \beta}  \ +  \sum_a (g_a^2 / 2) D_a^2 \ =  \ 3 m_{3/2}^2 M_P^2 \ , \label{soft8}
\end{equation}
where $\alpha, {\bar \beta}$ refers to fields contributing to
supersymmetry breaking and $a$ is an index for anomalous $U(1)$ gauge
factors. Then the most general formulae 
for soft terms of matter fields\footnote{We don't write the analytic
bilinear soft terms, since their discussion depends on the origin of
the corresponding ($\mu$-like) term in the superpotential.} 
$M^I$  ($F^I=0$), are given by \cite{Dudas:2005vv} (see also \cite{kawamura} for the heterotic strings case) 
 \begin{eqnarray}
&& m^2_{I {\bar J}} \  = \ m_{3/2}^2 \ K_{I {\bar J}} \ - \ F^{\alpha}
\
F^{\bar \beta} R_{{\alpha} {\bar \beta} I {\bar J}} \  - \  \sum_a g_a^2 D_a ( {1 \over 2} K_{I {\bar J}}
- \partial_I  {\partial_{\bar J}} ) D_a   \ , \nonumber \\
&& A_{IJK}  =  m_{3/2}^2 
\left( 3 \nabla_I \nabla_J G_K  +  G^{\alpha} \nabla_I \nabla_J
  \nabla_K G_{\alpha} \right)  
 -  g_a^2 D_a ( {D_a \over 2} \nabla_i \nabla_j G_k  -   \nabla_i
\nabla_j \nabla_k D_a) \ , \nonumber \\
&& M_{1/2}^a \ = \ {1 \over 2} (Re f_a)^{-1} \ m_{3/2} \ G^{\alpha} \
\partial_{\alpha} f_a \ , \label{soft9}
\end{eqnarray}
where $G = K + \ln |W|^2$, $G_{\alpha} = \partial_{\alpha} G$,
$\nabla_I G_J \  = \ G_{IJ} - \Gamma_{IJ}^K G_K$, etc., where 
$R_{{\alpha} {\bar \beta} I {\bar J}} \ =
\ \partial_{\alpha} \partial_{\bar \beta} \ K_{I {\bar J}} \ - \
\Gamma_{\alpha I}^M \ K_{M {\bar N}} \Gamma_{{\bar \beta} {\bar
J}}^{\bar N}$ is the Riemann tensor of the Kahler manifold and 
$\Gamma_{\alpha I}^M \ = \ K^{M {\bar N}} \partial_{\alpha} K_{{\bar
N} I}$ are the Christoffel symbols. Moreover,
\begin{equation}
D_a \ = \ X_I^a M^I \partial_I K  \ - \  {\eta_a^{\alpha}  \over 2} \partial_{\alpha} K \ . \label{soft10}  
\end{equation}
 In (\ref{soft10}), $X_I^a$ denote $U(1)_a$ charges of charged fields $M^I$ and
$\eta_a^{\alpha}$ are defined by the nonlinear gauge transformations of the
 moduli fields under (super)gauge fields transformations 
 \begin{equation}
 V_a \ \rightarrow V_a \ + \  \Lambda_a \ + \ {\bar \Lambda}_a \quad , \quad 
 T_{\alpha}  \ \rightarrow  \  T_{\alpha} \ + \    \eta_a^{\alpha}   \Lambda_a  \ . \label{soft11} 
 \end{equation}
 By using (\ref{soft10}), we can also write the scalar masses in (\ref{soft9}) as
  \begin{equation}
 m^2_{I {\bar J}}  \  = \ m_{3/2}^2 \ K_{I {\bar J}}  -  F^{\alpha} \
F^{\bar \beta} \ R_{{\alpha} {\bar \beta} I {\bar J}}  \  - \ \sum_a g_a^2 D_a ( {1 \over 2} D_a 
-  X_I^a -  v_{l} X_{ l}^a \partial_{ l}  \ +  \  {\eta_a^{\alpha} \over 2} \partial_{\alpha} ) 
 \ K_{I {\bar J}} \ , \label{soft12}
\end{equation}
where $v_{ l}$ are vev's of charged scalar fields $z^l$ of charge $X_l^a$.
An interesting question is  : In which simple cases the tree-level contributions of order $m_{3/2}$ in (\ref{soft12}) do cancel each other ? 
This question is particularly relevant in order to identify (classes of)  models in which loop contributions and in particular the 
anomaly-mediated contributions \cite{anomaly} are important. 
  
From a 4d point of view,  we are aware of three simple cases :

i) the well-known case of no-scale models \cite{noscale} , with $K_{T {\bar T}} |F^T|^2 = 3 m_{3/2}^2 M_P^2$,  $D_a=0$, with matter fields
having modular weights $n_I = - 1$ in (\ref{general01}), when $|F^T|^2 R_{T {\bar T} I {\bar J}} =
m_{3/2}^2 K_{I {\bar J}}$ . This generalizes easily to the case of several Kahler moduli $T_{\alpha}$. Starting from the effective lagrangian
\begin{equation}
K \ = \ - \sum_{\alpha} p_{\alpha} \ln (T_{\alpha} + {\bar T}_{\alpha}) \ + \ \prod_{\alpha}  (T_{\alpha} + {\bar T}_{\alpha})^{n_I^{\alpha}} |M^I|^2
\ + \cdots \ , \label{general1}  
\end{equation}
the no-scale structure is defined by the condition that the superpotential  $W $ is
 {\it independent }  of  $ T_{\alpha}$ and the (semi)positivity of the
scalar potential. Zero cosmological constant then implies
\begin{equation}
K^{\alpha} K_{\alpha} \equiv K_{\alpha \bar \beta} K^{\alpha} K^{\bar \beta} \ = \ 3 \quad \rightarrow \quad \sum_{\alpha} p_{\alpha} \ = \ 3 \ . \label{general2}  
\end{equation}
The condition of having tree-level zero soft scalar masses and A-terms 
for matter fields $M^I$ is then
\begin{equation}
\sum_{\alpha}  n_I^{\alpha} \ = \ -1  \ . \label{general3}
\end{equation} 

ii) When the following conditions are  simultaneously satisfied : \\
- D-term contributions are much  larger\footnote{We should keep in
  mind, however, that in supergravity with 
$\langle W \rangle \not=0$, there
is no pure D-breaking. This case assumes therefore $D_a \gg F^{\alpha}$, but F-terms have to exist.}  than the F-terms and cancel the cosmological constant $\sum_a (g_a^2/2) D_a^2 \simeq 
3 m_{3/2}^2$.  \\
-  there are no (large) vev's of charged scalar fields   $v_{l} = 0$.  \\
- the matter fields are neutral under the $U(1)$'s symmetries  and
come from the D3 brane sector  (or, more generally $n_I = - 1$) .

Indeed, in this case by using the Kahler potential 
 \begin{equation}
 K \ = \ - 3 \ \ln (T + {\bar T}) \ + \    (T + {\bar T})^{-1} \ |M^I|^2 \ + \  \cdots \ , \label{general4}
 \end{equation} 
then it can be easily checked that  the D-term contributions precisely
cancel the other terms in the soft terms in (\ref{soft9}). The
generalization of 
this D-dominated supersymmetry breaking case to the case of several moduli 
$T_{\alpha}$ is more involved and will not be discussed here.

iii) A simple way to obtain tree-level zero soft
masses is by geometric sequestering \cite{anomaly},  i.e separating in the internal space
the source of supersymmetry breaking from the matter fields. From a 4d viewpoint, the 
vanishing of the tree-level soft terms appear as non-trivial cancellations in the
general formula (\ref{soft9}). However this cancellation is protected
from quantum corrections by the geometric separation of the source of
supersymmetry breaking. A typical example, obtained by assuming
that moduli fields (in particular the modulus $T$) were stabilized in a supersymmetric way, is that of a matter field $M$ and a hidden sector
field $\phi_h$, which is the only source of supersymmetry breaking and of cancellation of the cosmological constant
$G_h G^h =3$. The 4d supergravity action is
\begin{eqnarray}
&& K \ = \ - 3 \ln \ ( 1 \ - \ {|M|^2 \over 3} \ - \ {|\phi_h|^2 \over 3}  ) \ , \ \nonumber \\
&& W \ = \ W_v (M) \ + \ W_h (\phi_h) \ . \label{general6} 
\end{eqnarray} 
 
It is also possible that a matter-like field $C$ with couplings to the observable matter saturates the vacuum energy  
$ K_{C {\bar C}} |F^C|^2 = 3 m_{3/2}^2 M_P^2$ and by fine-tuning provides the
cancellation of the tree-level soft scalar mass, 
see e.g. \cite{lnr}. 
When neither of these cases occur, other manifestly supersymmetric uplifting mechanism
are expected to  lead to soft scalar masses of the order of the gravitino mass  $m_{I {\bar J}}^2 \sim m_{3/2}^2$.

%%%%%%%%%%%%%%%%%%%%%%%%%%%%%%%%%%%%%%%%%%%%%%%%%%%%%%%%%%%%%%%%%%%%%
\subsection{Soft terms with dynamical  F-term uplifting} 

A particularly important question is the magnitude of the soft terms
in the visible sector in the present setup. In order to answer this
question, we first estimate the contribution to supersymmetry
breaking from the various fields. By using the results of section 2,
we find in the leading order
\begin{eqnarray}
&& \overline{F^{\varphi}} \ \equiv \ e^{K / 2} \ K^{\varphi {\bar
\varphi}} D_{\varphi} \ W \ \simeq \ e^{K / 2} \ K^{\varphi {\bar
\varphi}} \ ( {\bar \varphi}_0 W \ + \ \delta \Phi \
\partial_{\Phi}
\partial_{\varphi} W_2 ) \ \simeq \ 0 \ , \nonumber \\
&& \overline{F^{\tilde \varphi}} \ \simeq \ 0 \quad , \quad
\overline{F^{\Phi}} \ = \ e^{K / 2} \
\left(
\begin{array}{cc}
0  & 0
\\
0 &   \  - h
\mu^2 I_{N_f-N}  
\end{array}
\right)   \ , \nonumber \\
&& F^T \ \simeq \  \  {a \over (T_0 + {\bar T}_0)^{1 /2}} \ e^{-b T_0} \
\simeq \  \  - \ {3 \over b} \ m_{3/2} \ . \label{soft1}
\end{eqnarray}
Notice that the main contribution to supersymmetry breaking comes
from the magnetic mesonic fields $\Phi$, which are the main
responsible for the uplift of the vacuum energy
\begin{equation}
Tr (|F^{\Phi}|^2) \ \simeq \ 3 \ m_{3/2}^2 \ . \label{soft2}
\end{equation}
The transmission of supersymmetry breaking in the observable sector
depends on the couplings of the observable fields $M^I$ to the SUSY
breaking fields $\Phi$, $T$.  The relevant couplings for our present
discussion are the following terms in the Kahler metric of the matter fields $M^I$
\begin{equation}
 K_{I  {\bar J}} \ =  \ \ (T + {\bar T})^{n_I} \ Z_{I {\bar
J}} +  Tr (|\Phi|^2) \ Z'_{I {\bar J}} \ \ , \label{soft3}
\end{equation}
where the form of the $\Phi$ coupling in the Kahler metric in (\ref{soft3}) is dictated
by the diagonal $SU(N_f)$ flavor symmetry left unbroken by the mass
parameter $\mu$ in the ISS lagrangian.  The Yukawa couplings  $W_{IJK} $ could also depend on $T$ and
$\Phi$. 

Then from (\ref{soft9}) with  no D-term contributions $D_a=0$, we find 
that the $F^T$ contribution is subleading by  a
factor $1 / b^2 (T + {\bar T})^2$ with respect to the other
contributions. This has the nice feature that the 
flavor-dependent $F^T$ contribution to scalar soft masses are subleading. 
The result for the (canonically normalized scalars) soft masses, at the leading order, is then given by
\begin{eqnarray}
&& m^2_{I {\bar J}} \ = \ m_{3/2}^2 \ \delta_{I {\bar J}} \ + \ { h^2  (N_f-N) \
\mu^4 \over  (T + {\bar T})^3} \  (K^{-1} Z')_{I {\bar J}}  \ \
\nonumber \\
&& \simeq \ m_{3/2}^2 \left( \ \delta_{I {\bar J}} \ + \  3 
\  (K^{-1} Z')_{I {\bar J}} \ \right) \ .  \label{soft5}
\end{eqnarray}
If the coupling to the mesonic fields $\Phi$ is small, i.e the coefficients $Z'_{I {\bar J}}$ are
 suppressed, soft scalar masses in the observable
(MSSM) sector are universal and are similar with the ones obtained
in the " dilaton-dominated" scenario in the past. It would be very interesting to find physical reasons of why  
$Z'_{I {\bar J}}$ are small. The geometrical sequestering cannot be invoked in this case since the matter fields 
$M$ and the mesons $\Phi$ do not fit into the structure (\ref{general6}).       
If the coeff. $Z'_{I {\bar J}}$ are of order one, the two terms in (\ref{soft5}) are of the
same order and the flavor problem of gravity mediation is back.

A similar conclusion holds for the other possible source of flavor
violation, the A-terms. 
If the couplings of the mesons to the matter fields are small, we get
in the leading order, for the canonically normalized
scalars
\begin{equation}
A_{IJL} \ \simeq \ 3 \ m_{3/2} \ w_{IJL} \ , \label{soft06} 
\end{equation}
where $w_{IJL}$ are the low-energy Yukawa couplings  for the matter fields, related to the corresponding SUGRA 
couplings   $W_{IJL} = \nabla_I \nabla_J \nabla_L \ W$  by  
\begin{equation}
w_{IJL} = e^{K/2} \ (K^{-1/2})_I^{I'}   (K^{-1/2})_J^{J'}   (K^{-1/2})_L^{L'} \ W_{I'J'L'} \ . \label{soft010}
\end{equation}
Since A-terms are proportional
to the Yukawa couplings, there are no flavor violations in this case.    

Gaugino masses in the observable sector are determined by the gauge
kinetic functions which in our case have generically the form
\begin{equation}
f_a \ = \ f_a^{(0)} \ + \ \alpha_a T \ + \ \beta_a \ (Tr \Phi) \ ,
\label{soft6}
\end{equation}
where $f_a^{(0)}$ are provided by other moduli fields, stabilized in
a supersymmetric manner. The form of coupling to the mesons in
(\ref{soft6}) is fixed by the diagonal $SU(N_f)$ flavor symmetry
left unbroken by the mass parameter $\mu$, whereas $\alpha_a$ are numbers of
order one\footnote{In a type IIB orientifold embedding, this happens
if the observable sector lives on D7 branes.}. The gaugino masses
\begin{equation}
M_a \ =  \alpha_a F^T \ + \ \beta_a \ (Tr F^\Phi) \  \label{soft7}
\end{equation}
are of the order of the gravitino mass if $\beta_a$ are of order one, whereas they are supressed by the factor 
$1/ b (T + {\bar T})$ if $\beta_a$ are small. In this second case, the anomaly-mediated contributions 
\cite{anomaly,rattazzi} are comparable to the tree-level ones.
To conclude, we do not find a suppression of all of the soft terms in the observable sector
with respect to the gravitino mass. This is in
agreement with the results of ref. \cite{lnr}. 
Therefore our results point towards a gravity-mediation type
of supersymmetry breaking in the hidden sector, which in the case of
small couplings of matter to hiden sector mesons are very similar to
the dilaton-domination scenario and are therefore flavor blind at
tree-level \footnote{For other ways of getting flavor universality in
compactifications with stabilized moduli, see e.g. \cite{quevedo}.}

We would like to briefly compare these results to the ones
obtained in \cite{Choi:2004sx} by using the original KKLT uplifting
mechanism with D${\bar 3}$ antibranes\footnote{See also \cite{luty} for a model with a phenomenology similar to 
the one in \cite{Choi:2004sx}.}. By using a  nonlinear supergravity approach, 
 \cite{Choi:2004sx}  found a (moderate) hierarchy $m_{3/2} \sim 4 \pi^2 m_{soft}$.
Let us try to understand better  the difference with our results. 
As we discussed in the previous section, there are three ways of supressing the tree-level soft masses for matter fields.  
The first is no-scale type models.
The KKLT-type models are  not of this type, since $F^T$ contribution is small.
The second case is the dominant D-term breaking. 
This is probably the manifestly supersymmetric case which should correspond  
in the low energy limit to the analysis done in  \cite{Choi:2004sx}. Knowing that  pure  D-term supersymmetry
breaking does not exist, it could be  difficult  to realize a model
along these lines. It is however very interesting to investigate this
possibility in more detail.

We believe that a more  detailed phenomenological
analysis of the possible manifestly supersymmetric uplifting mechanisms  deserves further investigation.
%%%%%%%%%%%%%%%%%%%%%%%%%%%%%%%%%%%%%%%%%%%%%%%%%%%%%%%%%%%%%%%%%%%%%%%%%%%%%%%%%%%%%%%%%%%%%%%%%%%%%%%%%%%%%%%
\section*{Acknowledgments}{ We would like to thank  Z. Chacko, Z. Lalak, Y. Mambrini, A. Romagnoni,
C. Scrucca and R. Sundrum for useful discussions. E.D thanks KITP of Santa Barbara and 
S.P.  thanks the CERN theory group , respectively, for
hospitality during the completion of this work. Work partially
supported by the CNRS PICS \#~2530 and 3059, RTN contracts
MRTN-CT-2004-005104 and MRTN-CT-2004-503369, the European Union
Excellence Grant, MEXT-CT-2003-509661, by the Polish grant MEiN 1 P03B 099 29, the EC contract                                          
MTKD-CT-2005-029466  and by the National Science
Foundation under Grant No. PHY99-07949. }

%%%%%%%%%%%%%%%%%%%%%%%%%%%%%%%%%%%%%%%%%%%%%%%%%%%%%%%%%%%%%%%%%%%%%%%%%%%%%%%%%%%%%%%%%%%%%%%

\nocite{}
\bibliography{bmn}
\bibliographystyle{unsrt}

\end{document}